\documentclass[journal=jpcl,manuscript=letter]{achemso}

\usepackage{chemformula} 
\usepackage[T1]{fontenc} 

\usepackage{graphicx}
\usepackage{dcolumn}
\usepackage{textcomp}
\usepackage{color}
\usepackage[english]{babel}
\usepackage{amsmath}
\usepackage{amssymb}
\usepackage{bm}
\usepackage{upgreek}
\usepackage{overpic} 
\usepackage{siunitx}
\usepackage{xcolor}

\author{Damir Sakhapov}
\affiliation[Georg August]
{III. Institute of Physics -- Biophysics, Georg-August-University, 37077 G\"ottingen, Germany}
\author{Ingo Gregor}
\affiliation[Georg August]
{III. Institute of Physics -- Biophysics, Georg-August-University, 37077 G\"ottingen, Germany}
\author{Narain Karedla}
\affiliation{The~Rosalind~Franklin Institute, Harwell Campus, Didcot, OX11 0FA, UK}
\author{J\"org Enderlein}
\email{jenderl@gwdg.de}
\phone{+49 (0)551 3926908}
\fax{+49 (0)551 3927720}
\affiliation[Georg August]
{III. Institute of Physics -- Biophysics, Georg-August-University, 37077 G\"ottingen, Germany}
\alsoaffiliation[MBExC]
{Cluster of Excellence ``Multiscale Bioimaging: from Molecular Machines to Networks of Excitable Cells'' (MBExC), Georg August University, 37077 G\"ottingen, Germany}

\title[FCS photophysics]
  {Measuring photophysical transition rates with fluorescence correlation spectroscopy and antibunching}

\abbreviations{IR,NMR,UV}

\begin{document}

\begin{tocentry}
\begin{center}
\includegraphics[width=7cm]{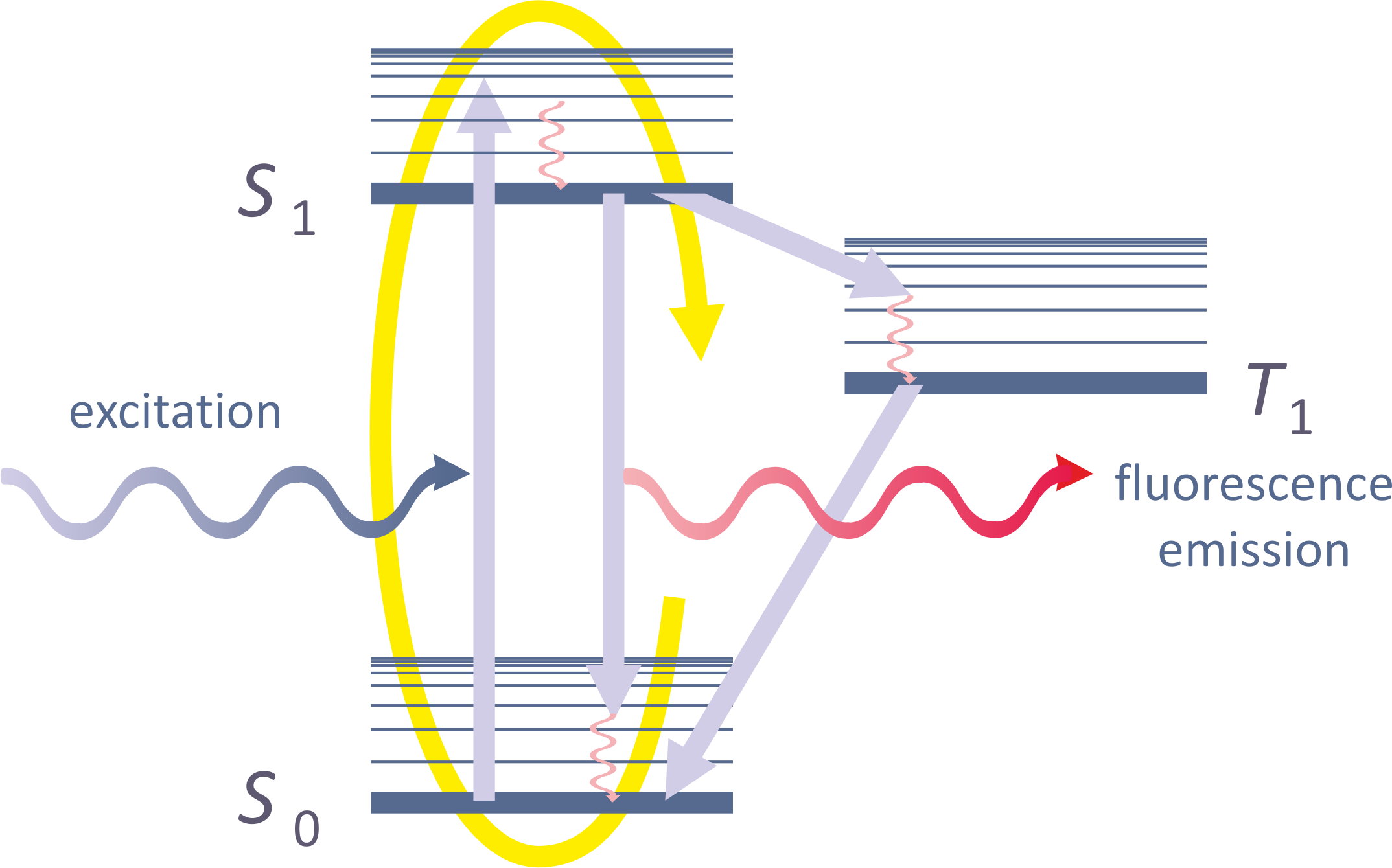}%
\end{center}
\end{tocentry}

\begin{abstract}
We present a new method that combines fluorescence correlation spectroscopy (FCS) on the microsecond time scale with fluorescence antibunching measurements on the nanosecond time scale for measuring photophysical rate constants of fluorescent molecules. The antibunching measurements allow us to quantify the average excitation rate of fluorescent molecules within the confocal detection volume of the FCS measurement setup. Knowledge of this value allows us then to quantify, in an absolute manner, the intersystem crossing rate and triplet state lifetime from the microsecond temporal decay of the FCS curves. We present a theoretical analysis of the method and estimate the maximum bias caused by the averaging of all quantities (excitation rate, photophysical rates) over the confocal detection volume, and we show that this bias is smaller than 5\% in most cases. We apply the method for measuring the photophysical rate constants of the widely used dyes Rhodamine~110 and ATTO~655.
\end{abstract}

\section{Introduction}

Fluorescent dyes have become indispensable for a myriad of microscopy and spectroscopy applications in the life sciences. Although the fundamental principles of fluorophore photophysics are already known since the start of quantum mechanics ca. 100 years ago, the quantitative measurement of the involved photophysical transition rates has become possible only with the advent flash- and laser photolysis and fast wide-band recording electronics~\citep{Lower_1965}. Newer techniques have used rapid fluid streams for kinetic measurements of dye photophysics \cite{Thiel1999b, Thiel2001}, but both photolysis or fluid stream techniques are experimentally challenging and require large amounts of sample. 

A powerful, single-molecule sensitive yet relatively simple technique for measuring fluorescence-related intensity fluctuations is fluorescence correlation spectroscopy (FCS), pioneered by Magde, Elson and Webb in 1972~\citep{MagdeFCS_1972}. This method allows for studying any dynamic process that modulates the fluorescence intensity measured within the tiny femtoliter-sized confocal volume of a confocal microscope. Initially, FCS was mostly used for determining diffusion coefficients of fluorescent molecules diffusing through the detection volume. Since the beginning of the nineties it was realized that FCS can be also used for determining fast photophysical transitions, in particular the intersystem crossing rate from the first excited state to the triplet state, and the transition rate from the triplet state to the ground state (i.e. triplet state lifetime)~\citep{widengren1994triplet, widengren1995fluorescence, widengren2000characterization, Widengren2008, Eggeling2008, blom2009triplet, schonle2014monitoring}. However, for obtaining absolute numbers for the photophysical transition rates, these FCS studies made independent estimates about the excitation intensity in the detection volume, which can result in considerable uncertainty for the final values.    

Already in 1997, Mets \textit{et al.} demonstrated that it is possible to directly determine absolute excitation rates of fluorescent molecules from fluorescence antibunching measurements \citep{rigler1997antibunch}. At that time, these measurements were quite demanding, with poor signal-to-noise ratio, and could not be readily combined with FCS experiments within the micro- to millisecond timescale. This was probably the reason for why their method was not picked up later by researchers who wanted to study the photophysics of fluorescent molecules with single-molecule sensitivity.  

In the present paper, we combine fluorescence antibunching measurements with FCS to measure absolute intersystem crossing rates and triplet state lifetimes of fluorescent molecules. As we will show below, this can be done within one single experiment where fluorescence intensity fluctuations are recorded with sufficiently high temporal resolution. Thus, both the excitation rate and the photophysical transition rates are simultaneously measured for exactly the same sample on the same experimental setup, which eliminates additional sources of error when one measures both these quantities in different experiments. In the next sections, we will develop the theoretical basis of our method, present numerical simulations for estimating its performance, and then experimentally apply it for measuring the photophysical transition rates of the widely used dyes Rhodamine~110 and ATTO~655.

\section{Theory}

\begin{figure}[bt]
    \centering
    \begin{overpic}[width=0.5\linewidth]{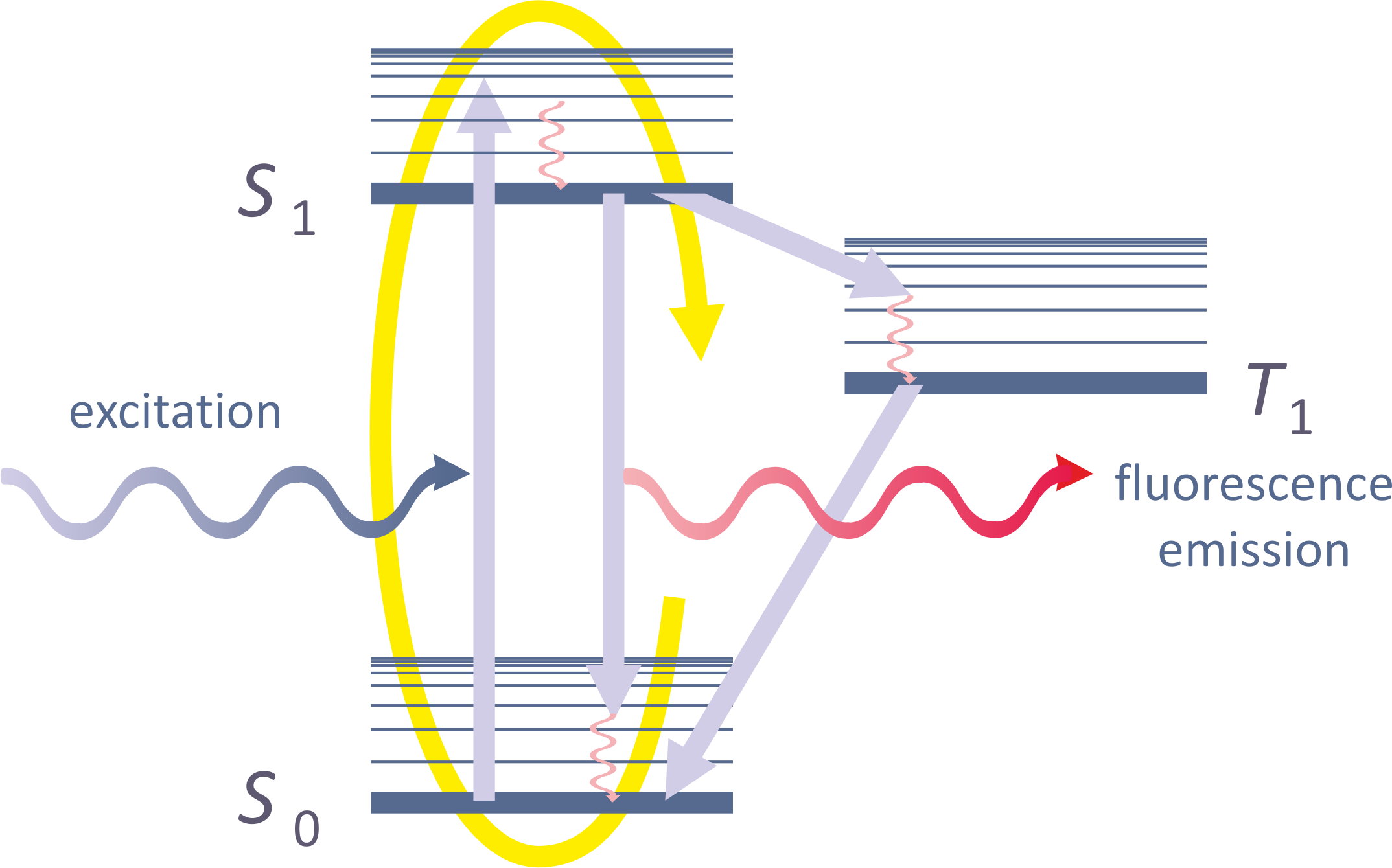}%
    \put(27.5,32) {$k_\mathrm{ex}$}%
    \put(46,32) {$\tau^{-1}$}%
    \put(56,47){$k_\mathrm{isc}$}%
    \put(56,14){$k_\mathrm{ph}$}%
    \end{overpic}
    \caption{Jablonski scheme of the considered photophysics: A fluorescent molecule has a singlet ground state ($S_0$) and first excited singlet state ($S_1$), and a triplet state $T_1$. The molecule is excited via single-photon excitation from $S_0$ into $S_1$, where it relaxes back into the lowest vibrational level within a few picoseconds. Subsequently, the molecule returns to the ground state $S_1$ with rate $\tau^{-1}$, where $\tau$ is the fluorescence lifetime (typically few nanoseconds), or it switches to the triplets state $T_1$ with intersystem crossing rate $k_\mathrm{isc}$, where in all relevant cases $k_\mathrm{isc}\ll\tau^{-1}$. From the triplet state, it eventually returns to the ground state with phosphorescence rate $k_\mathrm{ph}$.}
    \label{Fig1}
\end{figure}

We consider the three-level system shown in fig.~\ref{Fig1} consisting of an electronic singlet ground state $S_0$, first excited singlet state $S_1$, and a triplet state $T_1$. We will further assume that the transitions rate $k_\mathrm{fl}$ from $S_1$ to $S_0$ is by orders of magnitude faster than the intersystem crossing rate $k_\mathrm{isc}$ from state $S_1$ to $T_1$ or phosphorescence rate $k_\mathrm{ph}$ for the transition from $T_1$ to $S_0$. This allows us to consider the $S_0 \rightleftharpoons S_1$ photo-kinetics separately and decoupled from the photo-kinetics $\{S_0,S_1\} \rightleftharpoons T_1$.

Let us first consider the fast kinetics of transitions between the $S_0$ and $S_1$ states. If the molecule is in its ground state at time zero, the chance $s_1(t)$ to find it in its excited state $S_1$ at time $t$ is given by solving the kinetic equation
\begin{align}
\frac{ds_1(t)}{dt} = k_\mathrm{exc} - \left(k_\mathrm{exc}+\frac{1}{\tau}\right)s_1(t)
\label{eq:s1}
\end{align} 
where $k_\mathrm{exc}$ is the excitation rate that depends on the power density of the excitation laser, and where we have also taken into account that the chance to have the molecule in its ground state is $1-s_1(t)$. The solution of this equation reads 
\begin{align}
s_1(t) = \kappa\,\left\{1-\exp\left[-\left(k_\mathrm{exc}+\frac{1}{\tau}\right)t\right]\right\} \,\textrm{,} 
\label{eq:s1sol}
\end{align} 
where we have introduced the abbreviation 
\begin{align}
\kappa = \frac{k_\mathrm{exc}\tau}{1+k_\mathrm{exc}\tau}
\label{eq:kappa}
\end{align} 
which is the \emph{average excitation rate of a molecule} when it is \emph{not} in the triplet state, i.e. the limit for $t\rightarrow\infty$ of $s_1(t)$ in eq.~\eqref{eq:s1sol}. 

Next, let us consider the slow kinetics of intersystem crossing and phosphorescence, and let us assume that this takes place on such a slow time scale that at any moment in time, the fast transitions between $S_0$ and $S_1$ are in equilibrium. We denote the probability to find the molecule in any of its two singlet states by $s(t)$. Then, the probability of finding it in its first excited singlet state $S_1$ is given by $\kappa\,s(t)$, see also asymptotic value of solution \eqref{eq:s1sol}, and the time evolution of $s(t)$ is governed by the differential equation 
\begin{equation}
\begin{split}
\frac{ds(t)}{dt} &= - k_\mathrm{isc}\,\kappa\, s(t) + k_\mathrm{ph} \left[1-s(t)\right] \\
&= k_\mathrm{ph} - (k_\mathrm{ph} + \kappa\, k_\mathrm{isc})\,s(t)\,\textrm{.}
\end{split}
\label{eq:s}
\end{equation} 
With the initial condition $s(t=0)=1$, eq.~\eqref{eq:s} has the solution
\begin{equation}
\begin{split}
s(t) = \frac{k_\mathrm{ph}+\kappa k_\mathrm{isc}\exp\left[-(k_\mathrm{ph} + \kappa k_\mathrm{isc})t\right]}{k_\mathrm{ph}+\kappa k_\mathrm{isc}}
\end{split}
\label{eq:ssol}
\end{equation}

The autocorrelation function $g(t)$ measures the probability to detect a photon at time $(t_0+t)$ if there was a photon detection event at time $t_0$. The probability $p_0$ to detect a photon at $t_0$ (i.e. at an arbitrary time) is proportional to the probability that a molecule is \emph{not} in its triplet state (given by the asymptotic value of eq.~\eqref{eq:ssol} for large time $t$) times the probability that it is then in its excited state $S_1$ (given by $\kappa$ from eq.~\eqref{eq:kappa}). This has to be multiplied by the overall detection efficiency of the microscopy system $\epsilon$, so that we find
\begin{align}
p_0 = \frac{\epsilon\,\kappa\,k_\mathrm{ph}}{k_\mathrm{ph}+\kappa k_\mathrm{isc}}
\end{align} 
Directly after such an detection event the molecule had returned to its ground state, so that the probability to detect the \emph{next} photon is proportional to the probability $s_1(t)$, i.e. state occupancy of the first excited state. On a short timescale, this $s_1(t)$ is given by eq.~\eqref{eq:s1sol}, and on a longer timescale, it is given by the product of the probability $s(t)$ to find the molecule in one of its singlet states, see eq.~\eqref{eq:ssol}, times the probability $\kappa$ that it is then in its excited state, see eq.~\eqref{eq:kappa}. In a FCS experiment with a confocal microscope, see fig.~\ref{Fig2}, the situation is complicated by the fact that both the ratio $\kappa$ as well as the detection efficiency $\epsilon$ are functions of position $\mathbf{r}$ (the function $\kappa$ via the position-dependent excitation rate  $k_\mathrm{exc}$). When joining all these pieces together, the \emph{lag-time dependent} part of the correlation function reads (up to some constant factor)
\begin{equation}
\begin{split}
g(t) &= \int d\mathbf{r} p_0(\mathbf{r}) \epsilon(\mathbf{r}) s_1(t,\mathbf{r}) \\
&= \int d\mathbf{r} \frac{k_\mathrm{ph} \epsilon(\mathbf{r})^2 \kappa(\mathbf{r}) s_1(t,\mathbf{r})}{k_\mathrm{ph}+\kappa(\mathbf{r}) k_\mathrm{isc}} 
\end{split}
\label{eq:gphoto}
\end{equation}
where the integration extends over the whole sample volume. Thus, for lag-time values on the order of the fluorescence lifetime $\tau$, we have to use eq.~\eqref{eq:s1sol} for $s_1(t)$ and we explicitly find
\begin{align}
g_a(t) = \int d\mathbf{r} \frac{k_\mathrm{ph} \epsilon(\mathbf{r})^2 \kappa(\mathbf{r})^2}{k_\mathrm{ph}+\kappa(\mathbf{r}) k_\mathrm{isc}} \left\{1-e^{-\left[k_\mathrm{exc}(\mathbf{r})+\tau^{-1}\right]t}\right\}
\label{eq:ganti}
\end{align} 
which describes the well-known anti-correlation (antibunching) of the correlation function on very short timescales, and which is indicated by the index $a$ in $g_a$. For longer lag-time values on the timescale of the triplet-state lifetime $k_\mathrm{ph}^{-1}$, we have to use $s_1(t) = \kappa s(t)$ with $s(t)$ from eq.~\eqref{eq:ssol} and we find
\begin{equation}
\begin{split}
g_p(t) = \int d\mathbf{r} &\frac{k_\mathrm{ph} \epsilon(\mathbf{r})^2 \kappa(\mathbf{r})^2}{k_\mathrm{ph}+\kappa(\mathbf{r}) k_\mathrm{isc}} \\ &\cdot\frac{k_\mathrm{ph}+\kappa(\mathbf{r}) k_\mathrm{isc}e^{-(k_\mathrm{ph} + \kappa(\mathbf{r}) k_{\mathrm{isc}})t}}{k_\mathrm{ph}+\kappa(\mathbf{r}) k_\mathrm{isc}}
\end{split}
\end{equation}
where the index $p$ in $g_p$ indicates that this is the part of the correlation function that reflects the dye's photophysics. On even longer timescales, the temporal decay of the correlation function is governed by the diffusion of the molecules through the detection volume and will be modeled here by the conventional expression \citep{rigler2012fluorescence}
\begin{equation}
\begin{split}
g_d(t) = \frac{1}{(1+4Dt/a^2)\sqrt{1+4Dt/b^2}}
\end{split}
\label{eq:gdiff}
\end{equation}
where $D$ denotes the diffusion coefficient, $a$ and $b$ are the short and long axes of the assumed three-dimensional axially-symmetric Gaussian detection volume, and we have again omitted all constant factors. For sufficiently well-separated timescales of fluorescence lifetime, triplet-state photophysics, and diffusion (given by the characteristic diffusion time $a^2/4D$), the full correlation curve is well described by the product $g_a(t)g_p(t)g_d(t)$.  

Remarkably, experimentally measured correlation curves can be perfectly fitted with similar expressions but where all the position-dependent exponentials are replaced by single constant values, so that the short and long time parts of the correlation function are proportional to
\begin{align}
g_a(t) \propto 1-A e^{-\left(\langle k_\mathrm{exc}\rangle + \tau^{-1}\right)t}
\label{eq:gafinal}
\end{align} 
and
\begin{align}
g_p(t) \propto 1+T e^{-(k_\mathrm{ph} + \langle\kappa\rangle k_\mathrm{isc})t}
\label{eq:gpfinal}
\end{align} 
with constant factors $A$ and $T$, and with effective excitation rate $\langle k_\mathrm{exc}\rangle$ and effective \emph{mean} excitation rate $\left<\kappa\right>$. 

The core idea of our paper is now: (i) to use a fit of eq.~\eqref{eq:gafinal} to an experimentally measured antibunching curve for determining the effective excitation rate $\langle k_\mathrm{exc}\rangle$, provided that we know the fluorescence decay time $\tau$; (ii) to use this value $\langle k_\mathrm{exc}\rangle$ in eq.~\eqref{eq:kappa} for calculating an effective mean excitation rate $\left<\kappa\right>$; and (iii) to use this $\left<\kappa\right>$ to fit eq.~\eqref{eq:gpfinal} to an experimentally measured fluorescence correlation curve for extracting separate values for $k_\mathrm{ph}$ \emph{and} $k_\mathrm{isc}$. What has been done so far in the literature is to measure and fit correlation curves for different excitation intensities (i.e. values of $\left< k_\mathrm{exc}\right>$), to estimate the mean excitation rate $\left<\kappa\right>$ in the detection volume by \emph{independent measurements}, and then to fit the obtained dependence of $k_\mathrm{ph} + \langle\kappa\rangle k_\mathrm{isc}$ on $\left<\kappa\right>$ for determining the photophysical rate constants. Whereas the result for $k_\mathrm{ph}$ is relatively robust (it is the intercept of the curve $k_\mathrm{ph} + \langle\kappa\rangle k_\mathrm{isc}$ with the abscissa in the limit of zero excitation intensity), the value of $k_\mathrm{isc}$ crucially depends on the correct estimate of $\left<\kappa\right>$. Our idea here is to determine this value $\left<\kappa\right>$ in an intrinsic way from antibunching measurements. 

In the next section, we will check by numerical simulation what the expected bias and accuracy of this approach is, in the light that eqs.~\eqref{eq:gafinal} and \eqref{eq:gpfinal} are only rough approximations of the exact equations \eqref{eq:ganti} and \eqref{eq:gphoto}. In the section after the next section, we then apply our method for measuring the photophysical rate constants $k_\mathrm{ph}$ and $k_\mathrm{isc}$ of several dyes. 

\section{Numerical simulations} 

\begin{figure}[htb]
    \centering
    \includegraphics[width=0.4\linewidth]{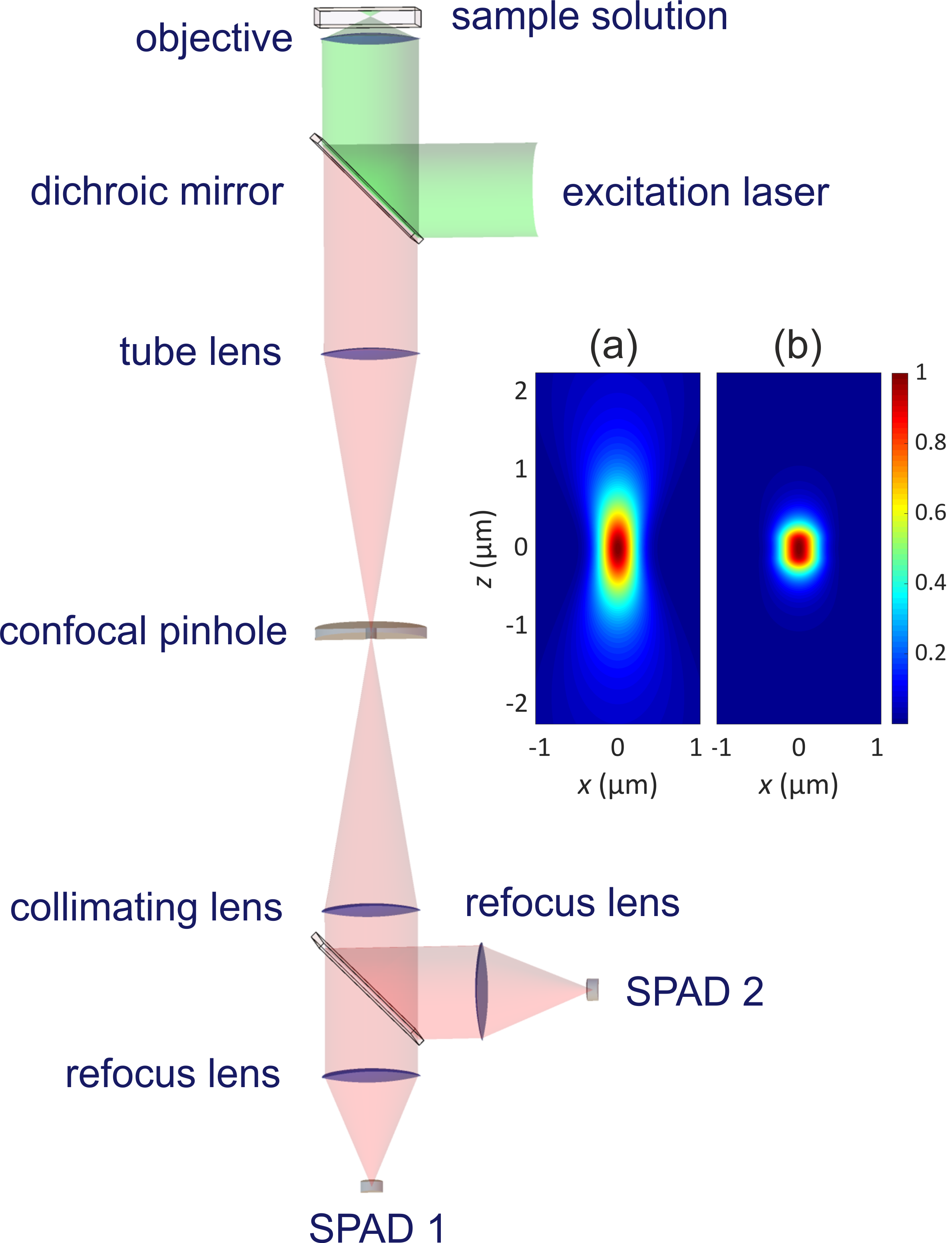}
    \caption{Schematic of the confocal setup used for all measurements. The inset \textbf{(a)} shows an exemplary excitation intensity distribution for a laser beam with 663~nm wavelength and focus beam waist of 350~nm, calculated by using eq.~\eqref{eq:excitation}, and inset \textbf{(b)} shows the full detection function $k_\mathrm{exc}\epsilon$ when detection is done through a pinhole with 25~µm radius and at 60$\times$ magnification, calculated using eq.~\eqref{eq:detectionefficieny}.}
    \label{Fig2}
\end{figure}

For simulating an FCS experiment, we use a simplified scalar theory of excitation and detection that was successfully used for modeling dual-focus fluorescence correlation experiments~\citep{Dertinger2007}. The excitation intensity distribution can be described as  scalar Gaussian laser beam \citep{Kogelnik_Li_1966}, which has been shown experimentally to be an excellent approximation of the light intensity distribution in the focus even when using high numerical aperture objectives~\citep{Benda_Benes_Marecek_Lhotsky_Hermens_Hof_2003, Machan_Hof_2010a,Machan_Hof_2010b}. The three-dimensional excitation rate distribution is given by
\begin{align}
k_\mathrm{exc}(x,y,z) = k_{\mathrm{exc},0} \exp\left[-\frac{2(x^2+y^2)}{w^2(z)}\right]
\label{eq:excitation}
\end{align} 
where $k_{\mathrm{exc},0}$ is the excitation rate at the center of the focus, $x$ and $y$ are the two Cartesian coordinates perpendicular to the optical axis, and $w(z)$ represents the beam diameter along the optical axis $z$ that evolves as~\citep{Kogelnik_Li_1966}
\begin{align}
w(z) = w_0 \sqrt{1+\left(\frac{\lambda_\mathrm{ex} z}{n\pi w_0^2}\right)^2}.
\label{eq:beamwaist}
\end{align}  
Here, $w_0$ is the beam waist in the focal plane $z=0$, $n$ denotes the refractive index of the sample, and $\lambda_\mathrm{ex}$ represents the vacuum wavelength of the excitation light. The detection efficiency $\epsilon$ can be approximated as~\citep{rigler1993fluorescence,qian1991analysis,Dertinger2007}
\begin{align}
\epsilon(z) = 1-\exp\left[-\frac{2a^2}{\rho(z)^2}\right]
\end{align}  
where $a$ is the radius of the confocal aperture divided by the magnification, and $\rho(z)$ is given by: 
\begin{align}
\rho(z) = \rho_0\sqrt{1+\left(\frac{\lambda_\mathrm{em}z}{n\pi \rho_0^2} \right)^2}
\label{eq:detectionefficieny}
\end{align}  
where $\lambda_\mathrm{em}$ is the vacuum wavelength of the emitted light, and $\rho_0 = \lambda_\mathrm{em}\pi/\tan\Theta$ is the radius of a diffraction-limited Gaussian-approximated intensity distribution in the focal plane that is obtained when focusing a plane wave through the microscope's objective, where $\Theta= \arcsin(\mathrm{N.A.}/n)$ is the half angle of the cone of objective's light collection with a numerical aperture N.A. This model of the detection function takes into account the attenuation of the light by the confocal pinhole only when an emitter is moved along the optical axis, but ignores any pinhole effects on the detection when the emitter is moved in the lateral plane. This approximation is acceptable as long as the excitation laser beam waist is smaller than the size of the pinhole. As an example, the two insets in Fig.~\ref{Fig2} show a $xz$-cross-section of the excitation intensity distribution $k_\mathrm{exc}$ (Fig.~\ref{Fig2}(a)) and the product $k_\mathrm{exc}\epsilon$ (Fig.~\ref{Fig2}(b)), for excitation wavelength $\lambda_{\mathrm{exc}}=663$~nm, a beam waist $w_0=350$~nm, and a pinhole radius divided by magnification of $a=25~$µm$/60=417$~nm.  

\begin{figure}[htb]
    \centering
    \includegraphics[width=0.95\linewidth]{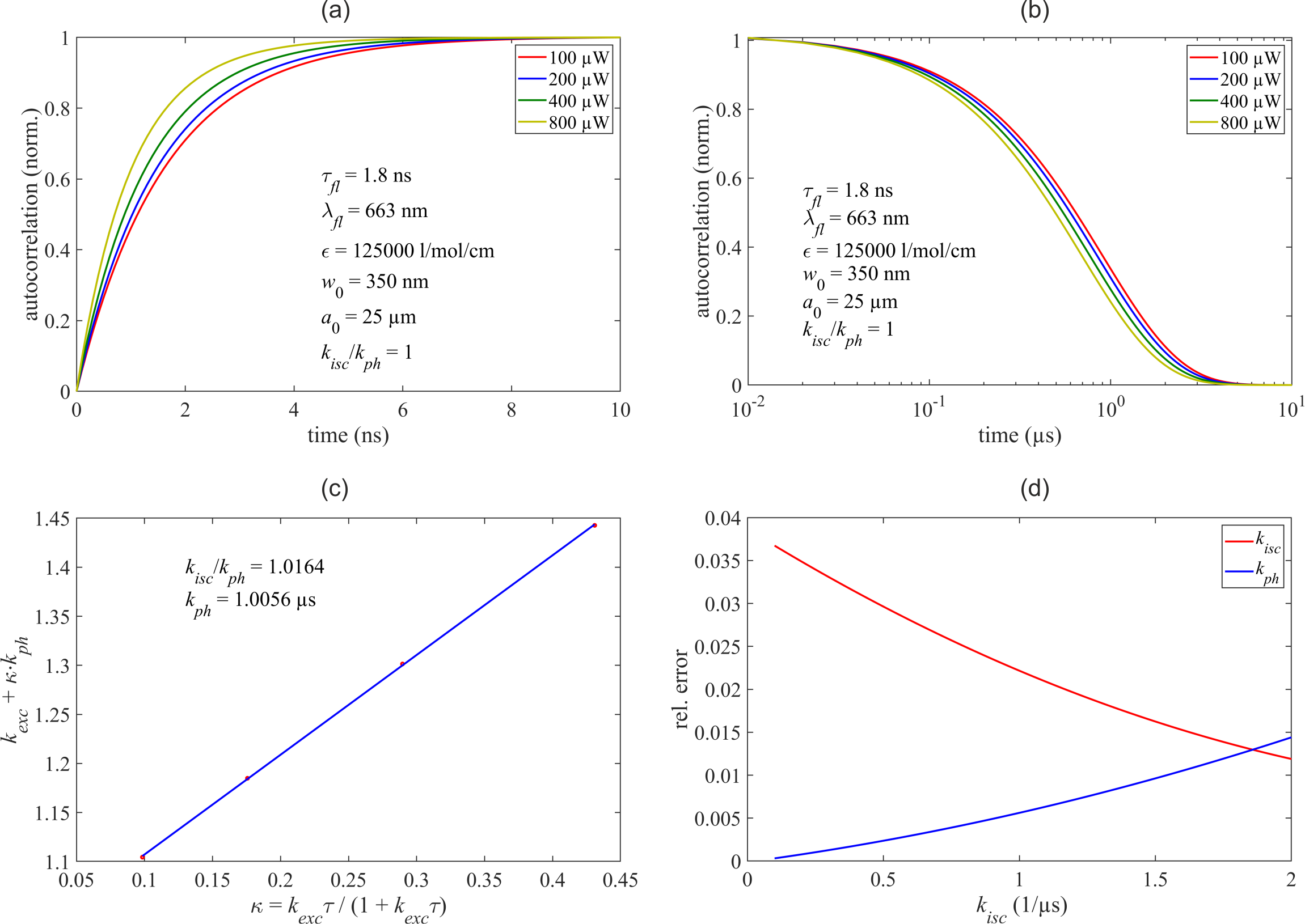}
    \caption{Model calculations. \textbf{(a)} Antibunching curves. \textbf{(b)} FCS curves for lag times longer than 10~ns. \textbf{(c)} Linear fit of $k_\mathrm{ph}+k_\mathrm{isc}\left<\kappa\right>$ against $\left<\kappa\right>$, where $k_\mathrm{ph}+k_\mathrm{isc}\left<\kappa\right>$ was obtained from fitting the correlation curves in (b), and $\left<\kappa\right>$ was obtained from fitting the antibunching curves in (a). The shown values for $k_\mathrm{isc}/k_\mathrm{ph}$ and $k_\mathrm{ph}$ are the estimates of the actual values obtained from the linear fit. \textbf{(d)} Relative errors for the estimates of $k_\mathrm{isc}$ and $k_\mathrm{ph}$ as a function of the actual value of $k_\mathrm{isc}$, with the value $k_\mathrm{ph}$ fixed to 1~µs$^{-1}$.}
    \label{Fig3}
\end{figure} 

Using the above model functions for $k_\mathrm{exc}(x,y,z)$ and $\kappa(z)$, we first calculated an antibunching curve using eq.~\eqref{eq:ganti}, fitted this curve with the mono-exponential fit function as written in eq.~\eqref{eq:gafinal} to obtain an ``averaged'' value of $\left<k_\mathrm{exc}\right>$, and thereafter used this value to calculate $\left<\kappa\right>=\left<k_\mathrm{exc}\right>\tau/(1+\left<k_\mathrm{exc}\right>\tau)$. Next, we calculated an FCS curve using eq.~\eqref{eq:gphoto}, fitted this curve with the mono-exponential expression in eq.\eqref{eq:gpfinal}, which yielded the exponential coefficient $k_\mathrm{ph}+k_\mathrm{isc}\left<\kappa\right>$. We repeated this procedure for several excitation intensities $I_0$ yielding the linear relation between 
$k_\mathrm{ph}+k_\mathrm{isc}\left<\kappa\right>$ and  $\left<\kappa\right>$. Using this linear relationship, estimates for $k_\mathrm{ph}$ and $k_\mathrm{isc}$ were obtained that can be then compared against the actual set values used for the model calculations.  For all these calculations, we fixed excitation wavelength $\lambda_\mathrm{ex}=663$~nm, emission wavelength $\lambda_\mathrm{ex}=660$~nm, molar extinction coefficient of the molecule $\sigma = 125000$~l$\cdot$M$^{-1}\cdot$cm$^{-1}$, fluorescence lifetime $\tau=1.8$~ns, excitation laser beam waist $w_0 = 350$~nm, and intersystem crossing rate $k_\mathrm{isc} = 1$~µs$^{-1}$. The calculations were done for four different values of 100~µW, 200~µW, 400~µW, and 800~µW of excitation laser power $P$, where the corresponding excitation rates $k_{\mathrm{exc},0}$ (excitations per second) were calculated by
\begin{align}
k_{\mathrm{exc},0} = \frac{2 P}{\pi w_0^2} \frac{10^3 \ln(10) \sigma}{N_\mathrm{A}} \frac{\lambda_\mathrm{ex}}{h c }
\end{align}  
where $h$ is Planck's constant, $c$ the vacuum speed of light, and $N_\mathrm{A}$ the Avogadro-Loschmidt number. Fig.~\ref{Fig3}(a) shows the model results for the antibunching curves, and fig.~\ref{Fig3}(b) for the corresponding FCS curves. The integrals in eqs.~\eqref{eq:ganti} and \eqref{eq:gphoto} where approximated by summation over a discretized grid covering a cylinder with 2.1~µm radius and $\pm$2.8~µm axial extent, with grid spacing of 21~nm radially and 28~nm axially. In all the calculations so far the phosphorescence rate was set to $k_\mathrm{ph}=1$~µs$^{-1}$. The linear fit of $k_\mathrm{ph}+k_\mathrm{isc}\left<\kappa\right>$ as a function of $\left<\kappa\right>$ is shown in Fig.~\ref{Fig3}(c). Next, we repeated these calculations for a range of phosphorescence rates between 0.1~µs$^{-1}$ and 10~µs$^{-1}$, and the relative errors $k_\mathrm{estimate}/k_\mathrm{actual}-1$ of the estimated rate values for $k_\mathrm{isc}$ and $k_\mathrm{ph}$ are shown in Fig.~\ref{Fig3}(d). 

It is noteworthy that the relative errors, for both the intersystem crossing rate as well as phosphorescence rate, are below 4\% across the whole range of phosphorescence rates considered. It should also be noted that only the ratio $k_\mathrm{ph}/k_\mathrm{isc}$ is important, not the absolute rate values, so that our error estimate will be valid for all ratios $k_\mathrm{ph}/k_\mathrm{isc}$ considered here, independent on the absolute values. Thus, our model calculations show that despite the highly non-uniform distributions of $k_\mathrm{exc}$ and $\epsilon$ across the detection volume, fitting the resulting antibunching and FCS curves with mono-exponential functions, and then using these fits for estimating the photo-physical rates $k_\mathrm{isc}$ and $k_\mathrm{ph}$, delivers surprisingly accurate results within a relative error of 4\% (within the considered range of 0.1--10 for the ratio $k_\mathrm{isc}/k_\mathrm{ph}$). In the next sections, we will apply our rate estimation scheme to determining intersystem crossing and phosphorescence rates of several common fluorescent dyes.

\section{Experimental results}

We have measured the photophysics of two dyes, Rhodamine~110 in the green spectral region, and ATTO~655 in the red spectral region. We chose Rhodamine~110 because there exist already excellent publications that report its intersystem crossing and triplet-to-ground state transition rates~\citep{Eggeling2008}, against which we can validate the current method. ATTO~655 because is routinely used in photo-electron transfer spectroscopy for monitoring conformational dynamics in biomolecules (peptides, DNA). We benchmark the photophysical properties of this dye in this work which will is essential for separating the timescales of intensity fluctuations arising from its own photophysical transitions from the conformational dynamics of biomolecules.

\begin{figure}
    \centering
    \includegraphics[width=0.475\linewidth]{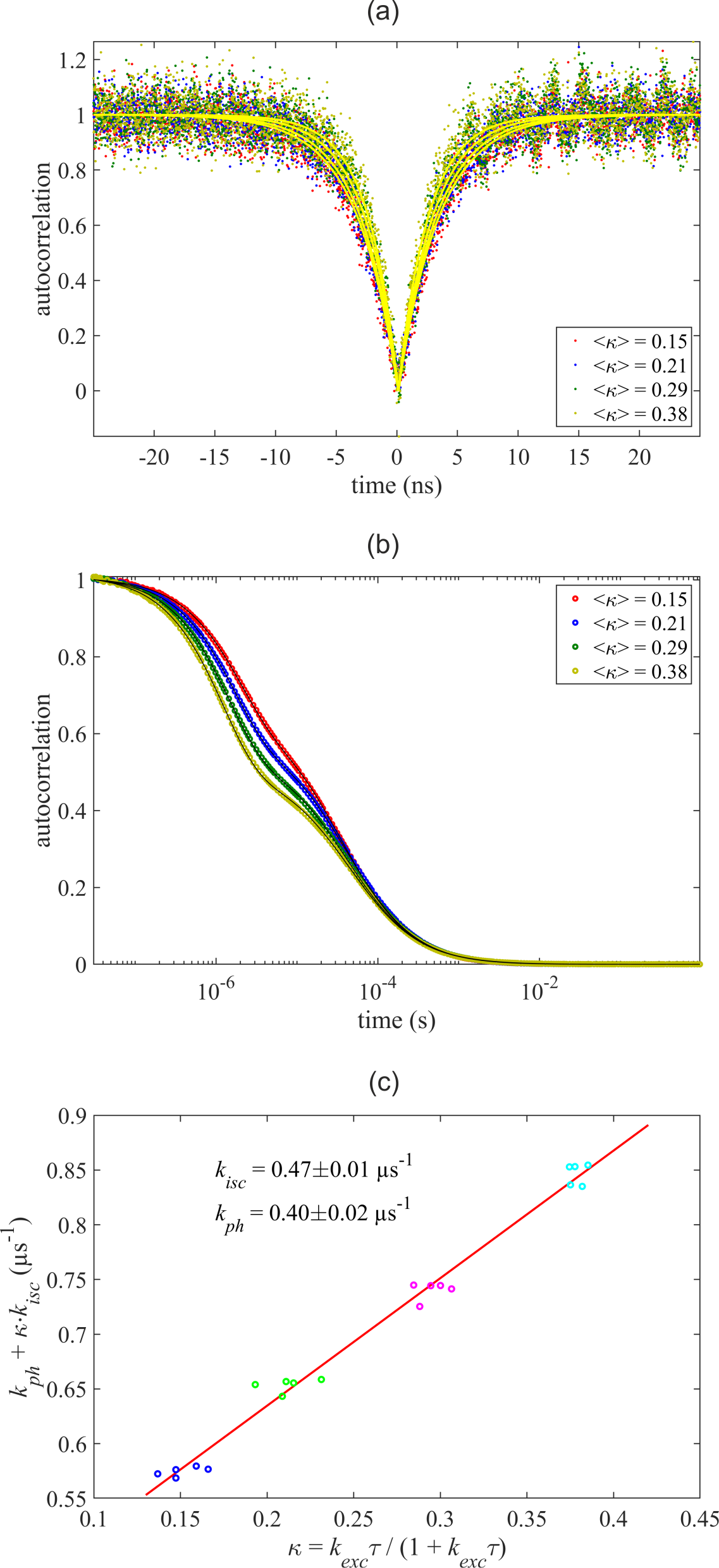}
    \caption{Photophysics measurements of Rhodamine~110. \textbf{(a)} Measured antibunching curves and fits (solid lines) for four different excitation laser powers. \textbf{(b)} Measured µs/ms-FCS curves with fits  (solid lines) for the same four different excitation powers as in (a). \textbf{(c)} Linear fit of the rate $k_\mathrm{ph}+k_\mathrm{isc}\left<\kappa\right>$ as extracted from the fits in (b) against $\left<\kappa\right>$ as extracted from the fits in (a). For each excitation power, the measurement was repeated five times, and the linear fit was repeated a thousand times, each time randomly choosing one of the five measurements for the four excitation powers. This yields the shown uncertainty values for $k_\mathrm{isc}$ and $k_\mathrm{ph}$ (standard deviation of the fitted value distributions).}
    \label{Fig4}
\end{figure}

For each dye, we performed four antibunching/FCS measurements at four increasing excitation intensity values (for measurement details see sections below). The results for Rhodamine~110 are shown in Fig.~\ref{Fig4}. The four (normalized) antibunching curves and the corresponding microsecond/millisecond FCS autocorrelation curves are shown in panels (a) and (b), respectively. We measured the fluorescence lifetime of Rhodamine~110 in a separate Time-Correlated Single-Photon Counting (TCSPC) measurement and obtained a fluorescence lifetime value $\tau=4.0$~ns which in perfect agreement with reported values. By fitting the antibunching curves with eq.~\eqref{eq:gafinal}, we obtain an estimate for $\left<k_\mathrm{exc}\right> + \tau^{-1}$, which allows us to determine $\left<k_\mathrm{exc}\right>$ and then subsequently to estimate $\left<\kappa\right> = \left<k_\mathrm{exc}\right>\tau/(1+\left<k_\mathrm{exc}\right>\tau)$. The microsecond decays of the autocorrelation curves shown in Fig.~\ref{Fig4}(b) are fitted with eq.~\eqref{eq:gpfinal}, again using a single exponential rise function where the exponential constant is equal to $k_\mathrm{isc} \left<\kappa\right> + k_\mathrm{ph}$. 

To obtain error estimates for our fit results, we applied a bootstrap procedure. We divided each measurement into ~10 sub-measurements, and then repeated the fitting five times using each time a randomly chosen set containing half of these sub-measurements. The resulting spread of fit results is visualized by the point clouds shown in Fig.~\ref{Fig4}(c). Then, we fitted the abscissa values $k_\mathrm{isc} \left<\kappa\right> + k_\mathrm{ph}$ linearly against the ordinate values $\left<\kappa\right> = \left<k_\mathrm{exc}\right>\tau/(1+\left<k_\mathrm{exc}\right>\tau)$, again using a bootstrap approach. A linear fit was performed on a randomly chosen set of four points, one from each point cloud. This was repeated $10^3$ times, and the resulting mean value and variance are textually shown in the figure. For Rhodamine~110, our methods yields an intersystem crossing rate of $k_\mathrm{isc}=0.47\pm0.01$~µs$^{-1}$, and a triplet-to-ground-state transition rate of $k_\mathrm{ph}=0.4\pm0.02$~µs$^{-1}$.

\begin{figure}
    \centering
    \includegraphics[width=0.475\linewidth]{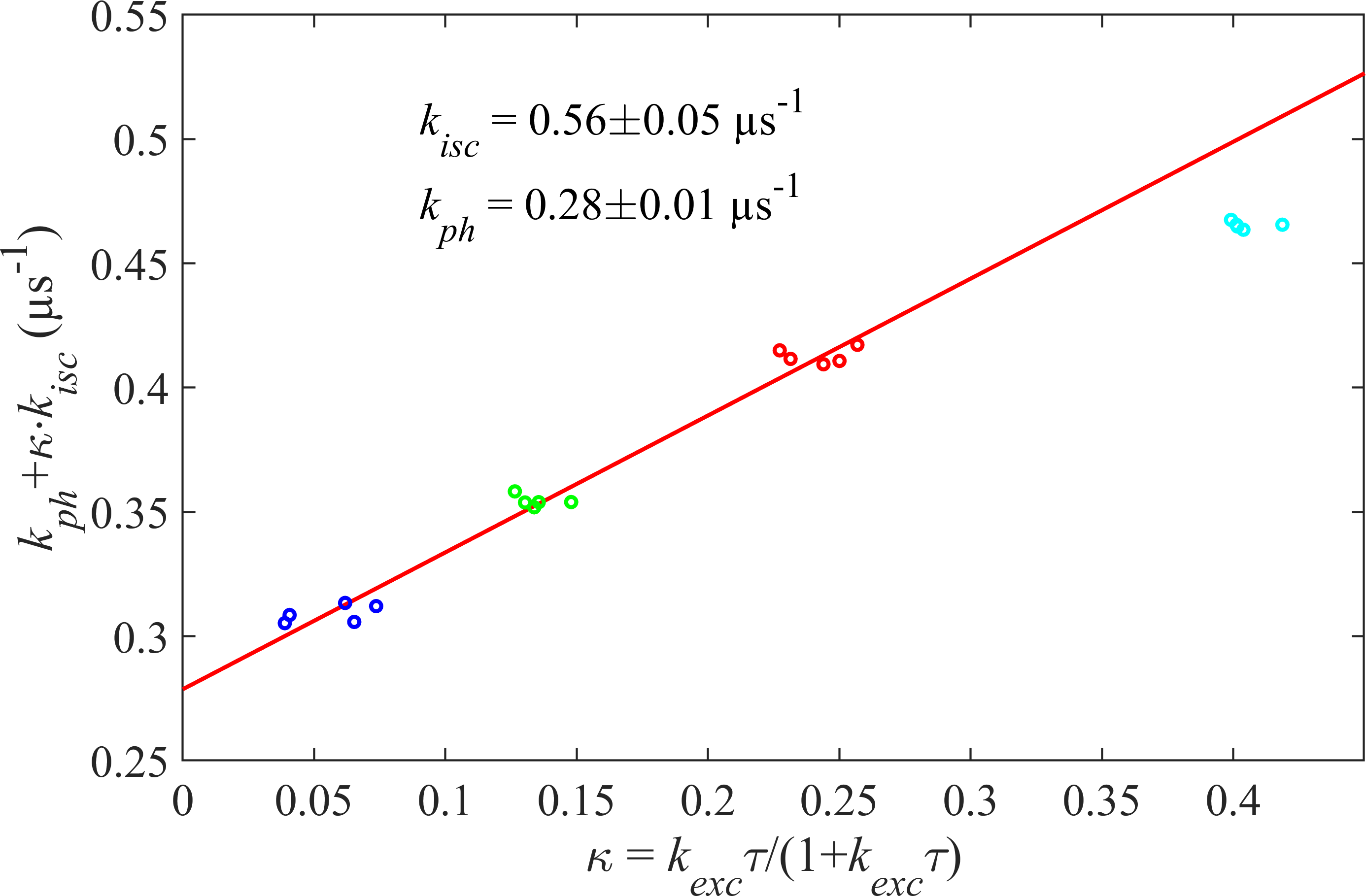}
    \caption{Photophysics of ATTO~655. Linear fit of $\left<\kappa\right>$ against $k_\mathrm{ph}+k_\mathrm{isc}\left<\kappa\right>$ for the three lowest-excitation measurements (three left point clouds). The highest excitation measurement (right-most point cloud) was excluded because we observe an obvious deviation from a linear relationship that we attribute to the activation of other photophysical channels at higher excitation power that are not included in our simple three-state model shown in Fig.~\ref{Fig1}.}
    \label{Fig5}
\end{figure}

The corresponding measurements and fit results for $\left<\kappa\right> = \left<k_\mathrm{exc}\right>\tau/(1+\left<k_\mathrm{exc}\right>\tau)$ and  $k_\mathrm{isc} \left<\kappa\right> + k_\mathrm{ph}$ for the dye ATTO~655 are shown in Fig.~\ref{Fig5}, where we have used the same fitting and bootstrap procedures for estimating the mean values and fit errors. For ATTO~655, we observe a clear deviation from linearity for the highest excitation intensity used, which we attribute to additional photophysical channels (e.g. triplet-to-singlet state back pumping) that are activated at higher excitation intensities and are not taken into account by our simple three-state-one-way model shown in Fig.~\ref{Fig1}. Thus, for the linear fit of $\left<\kappa\right>$ against $k_\mathrm{isc} \left<\kappa\right> + k_\mathrm{ph}$, we used only the three measurements corresponding to the three lowest excitation intensities. This gave us an intersystem crossing rate of $k_\mathrm{isc}=0.56\pm0.05$~µs$^{-1}$, and a triplet-to-ground-state transition rate of $k_\mathrm{ph}=0.28\pm0.01$~µs$^{-1}$.

Finally, one has to add to the experimental uncertainties of $k_\mathrm{isc}$ and $k_\mathrm{ph}$ as determined with our bootstrap analysis a potential systematic bias of below 4\% as indicated by our theoretical-numerical analysis in the preceding chapter (see Fig.~\ref{Fig3}), which amounts to a potential difference between our fit results and the actual values of d$\approx0.01-0.02$~µs$^{-1}$ for our transition rates of Rhodamine~110 and ATTO~655.

\section{Methods}
\subsection{Sample preparation}
Fresh samples were prepared and measured at room temperature. Both ATTO~655 and Rhodamine~110 dyes were diluted in PBS (pH 7.4). The final concentration of ATTO~655 was approximately 0.5 nM, while for Rhodamine~110 the concentration was 2 nM.
Experiments were always performed in 8-well chamber slides containing a 150~µl aliquot of freely diffusing dye solution sealed with parafilm to prevent evaporation during the measurements.Each intensity point was measured for 4 hours with the number of photons which depends on the excitation intensity with the highest value of 1.226 billion photons for ATTO~655 at the highest excitation intensity and 420 million photons at the lowest intensity point for the same dye. For Rhodamine~110, the number of photons ranges from 4 to 2 billions depending on the intensity point.

\subsection{Confocal Setup and measurements}

FCS measurements were performed at a commercially available MicroTime~200 (PicoQuant GmbH, Berlin, Germany) equipped with a water immersion objective (60x UplanSApo, NA 1.2, Olympus, Japan). All measurements were carried out with continuous-wave excitation. 488 nm laser (PhoxX 488-60, Omicron-Laserage Laserprodukte GmbH, Rodgau, Germany) was used to excite Rhodamine~110, whereas for ATTO~655 a 640 nm laser (LDH-D-C-640s, PicoQuant, Berlin, Germany) was utilized. The lasers were coupled into a single-mode fiber and then re-collimated using an apochromat lens. Thereafter, the laser beam was guided into the microscope with the help of a dichroic mirror. The fluorescence emission was collected by the same excitation objective, focused through a 50 um pinhole and focused onto two single-photon silicon avalanche photodiodes (SPCM CD 3516 H, Excelitas Technologies, Inc., Dumberry, Canada) after being split equally using a 50:50 beamsplitter. Additionally, appropriate bandpass filters were used to block any back-reflected laser (692/40 for ATTO~655 measurements and 525/45 for Rhodamine~110 measurements). In this series of measurements, every subsequent intensity point was set to be approximately twice higher than the previous, with a total laser power of up to 1~mW. Detected fluorescence photons were registered with the high-speed timing electronics HydraHarp 400 and SymPhoTime software (PicoQuant GmbH, Berlin, Germany) in time-tagged time-resoled mode. Raw data was correlated and fitted using custom written MATLAB (The Mathworks, Inc. Nattick, MA, USA) routines .

\section{Conclusion}
In this paper, we combined nanosecond FCS measurements (fluorescence antibunching) for estimating the absolute excitation intensity with micro- to millisecond FCS measurements for monitoring photophysical transitions (intersystem crossing, phosphorescence) to obtain calibration-free values for the photophysical transition rate constants. As our numerical simulations show, the systematic bias of the procedure is estimated to be smaller than $\sim$5\% for a wide range of rate constants. For the highest excitation intensity used int he ATTO~655 experiments, we observe a slight deviation of the expected linear relation between (effective) excitation rate and (effective) photophysical rate constants which hint at the onset of other photophysical processes that can be neglected at lower intensities. Thus, future generalizations of our method could extend the theoretical model underlying our data analysis for including such additional transitions and to determine their corresponding rate constants. An any case, we believe that our method presented here is a fast and satisfactorily precise method for determining photophysical rate constants on minute amounts sample.  

\section{Acknowledgments}
DS and JE acknowledge financial support by the Deutsche Forschungsgemeinschaft (DFG, German Research Foundation) via project A02 of the Collective research Collaboration SFB1456. JE acknowledges financing by the Deutsche Forschungsgemeinschaft (DFG, German Research Foundation) under Germany’s Excellence Strategy - EXC 2067/1- 390729940, and financing by the European Research Council (ERC) via project “smMIET” (Grant agreement No. 884488) under the European Union’s Horizon 2020 research and innovation programme. N.K. acknowledges EPSRC funding for the Rosalind Franklin Institute (RFI).

\providecommand{\latin}[1]{#1}
\makeatletter
\providecommand{\doi}
  {\begingroup\let\do\@makeother\dospecials
  \catcode`\{=1 \catcode`\}=2 \doi@aux}
\providecommand{\doi@aux}[1]{\endgroup\texttt{#1}}
\makeatother
\providecommand*\mcitethebibliography{\thebibliography}
\csname @ifundefined\endcsname{endmcitethebibliography}
  {\let\endmcitethebibliography\endthebibliography}{}

\end{document}